\documentclass[aps,prl,twocolumn,groupedaddress]{revtex4}
\usepackage{amsmath,amssymb,amsthm,mathrsfs,amsfonts,dsfont} 
\usepackage{graphicx}
\usepackage{subfigure}
\usepackage{amsbsy}
\usepackage{bm}
\usepackage{physics}

\usepackage{xcolor}
\newcommand{\blue}[1]{{\color{black}#1}}

\def\id{\mathds{1}}

\begin{document}

\title{Spin glass model of in-context learning}
\author{Yuhao Li$^{1}$}
\author{Ruoran Bai$^{1}$}
\author{Haiping Huang$^{1,2}$}
\email{huanghp7@mail.sysu.edu.cn}
\affiliation{$^{1}$PMI Lab, School of Physics,
Sun Yat-sen University, Guangzhou 510275, People's Republic of China}
\affiliation{$^{2}$Guangdong Provincial Key Laboratory of Magnetoelectric Physics and Devices,
Sun Yat-sen University, Guangzhou 510275, People's Republic of China}
\date{\today}

\begin{abstract}
	Large language models show a surprising in-context learning ability---being able to use a prompt to form a prediction for a query, yet without additional training, in stark
	contrast to old-fashioned supervised learning. Providing a mechanistic interpretation and linking the empirical phenomenon to physics are thus challenging and remain unsolved. We study a simple yet expressive transformer with linear attention and map this structure to a spin glass model with real-valued spins, where the couplings and
	 fields explain the intrinsic disorder in data. The spin glass model explains how the weight parameters interact with each other during pre-training, and further clarifies why an unseen function can be predicted by providing only a prompt yet without further training. Our theory reveals that for single-instance learning, increasing the task diversity leads to the emergence of in-context learning, by allowing the Boltzmann distribution to converge to a unique correct solution of weight parameters.  Therefore the pre-trained transformer displays a prediction power in a novel prompt setting. \blue{The proposed analytically tractable model thus offers a promising avenue for thinking about how to interpret many intriguing but puzzling properties of large language models.}   
\end{abstract}

 \maketitle

\textit{Introduction.}---
Thanks to earlier breakthroughs in processing natural languages (e.g., translation), vector representation and attention concepts were introduced into machine learning~\cite{Bengio-2003,WB-2013,Cho-2014,Bah-2015}, which further inspired a recent breakthrough of implementing the self-attention as a feedforward model of information flow, namely transformer~\cite{Att-2017}. The self-attention captures dependencies between different parts of the input (e.g., image or text), coupled with a simple cost of next-token prediction~\cite{Uni-2023,Li-2024}, leading to a revolution in the field of natural language processing~\cite{Sparks-2023}, so-called large language model (LLM).
 
 One of the astonishing abilities of the transformer is the in-context learning~\cite{LLM-2020}, i.e., the pre-trained transformer is able to accomplish previously-unseen complicated tasks by showing a short prompt in the form of instructions and a handful of demonstrations, especially without a need for updating the model parameters. LLMs thus develop a wide range of abilities and skills (e.g., question answering, code generation)~\cite{Radford-2019}, which are not explicitly contained in the training dataset and are not specially designed to optimize.
This remarkable property is achieved only by training for forecasting the next tokens and only if corpus and model sizes are scaled up to a huge number~\cite{Kaplan-2020,Wei-2022}. The above characteristics of transformer and the in-context learning (ICL) are in stark contrast to perceptron models in the standard supervised learning context, presenting a formidable challenge for a mechanistic interpretation~\cite{Persp-2023,Huang-2024}. 

To achieve a scientific theory of ICL, previous works focused on optimization via gradient descent dynamics~\cite{Von-2023,Birth-2023}, representation capacity~\cite{Garg-2022}, Bayesian inference~\cite{Xie-2021,Ganguli-2024}, and in particular the pre-training task diversity~\cite{Li-2023,Ganguli-2024,CP-2024,Wu-2024}. 
The theoretical efforts were commonly based on a single-layer linear attention~\cite{Aky-2022,Von-2023,CP-2024,Zhang-2024}, which revealed that a sufficient pre-training task diversity guarantees the emergence of ICL, i.e., the model can generalize beyond the scope of pre-training tasks. 

However, rare connections are established to physics models, which makes a physics model of ICL lacking so far, preventing us from a deep understanding of how ICL emerges from pre-trained model parameters. Here, we treat the transformer learning as a statistical inference problem, and then rephrase the inference problem as a spin glass model, where the transformer parameters are turned into real-valued spins, and the input sequences act as a quenched disorder, which makes the spins strongly interact with each other to lower down the ICL error. A unique spin solution exists in the model, guaranteeing that the transformer can predict the unknown function embedded in test prompts.
The derived formulas specify the intelligence boundary of ICL and how this can be achieved.

\textit{Transformer with linear attention.}---
We consider a simple transformer structure---a single-layer self-attention transforming an input sequence to an output one. Given an input sequence $\mathbf{X}\in\mathbb{R}^{D\times N}$, where $D$ is the embedding dimension and $N$ is the context length, the self-attention matrix is a softmax function $\text{Softmax} (\mathbf{Q}^\top\mathbf{K} / \sqrt{D})$, where $\mathbf{Q}=\mathbf{W}_{\text{Q}}\mathbf{X}$, $\mathbf{K}=\mathbf{W}_{\text{K}}\mathbf{X}$. $\mathbf{W}_{\rm Q}$ and $\mathbf{W}_{\rm K}$ are the query and key matrices ($\in\mathbb{R}^{D\times D}$), respectively.  $\sqrt{D}$ inside the softmax function makes its argument order of unity. The self-attention refers to the attention matrix generated from the input sequence itself and allows each element (query) to attend to all other elements in one input sequence, being learnable through pre-training. The softmax function is thus calculated independently for each row. Taking an additional transformation $\mathbf{V}=\mathbf{W}_{\text{V}}\mathbf{X}$, where $\mathbf{W}_{\rm V}\in\mathbb{R}^{D\times D}$ is the value matrix, one can generate the output $\mathbf{Y}=\mathbf{V}\cdot \text{Softmax} (\mathbf{Q}^\top\mathbf{K} / \sqrt{D})$. Hence, this simple transformer implements a function $\varphi_{\rm TF}(\mathbf{X}):\mathbb{R}^{D\times N}\to\mathbb{R}^{D\times N}$. 

For simplicity, we replace the computationally expensive softmax with linear attention,  which is \textit{still expressive}~\cite{Ahn-2024}. Defining $\mathbf{W} \equiv \mathbf{W}_{\text{Q}}^\top \mathbf{W}_{\text{K}}$, and choosing $\mathbf{W}_{\text{V}} = \id_D$  ($\id_D$ indicates a $D\times D$ identity matrix) for our focus on the query and key matrices, we re-express the linear transformer as \blue{$\mathbf{Y}=N^{-1}\mathbf{X}\mathbf{X}^\top \mathbf{W} \mathbf{X}$}, where $\mathbf{X}$ contains prompts and the query (to be predicted by the transformer), and $\mathbf{W} \in \mathbb{R}^{D \times D}$ is the equivalent weight matrix to be trained, and \blue{$N^{-1}$ is a normalization coefficient (see more explanations in SM~\cite{SM} which also discusses a separate training of query and key matrices}). 

We next design the training task as a high-dimensional linear regression. Each example consists of the data $\bm{x} \sim \mathcal{N}(0,\id_D)$ and the corresponding label $y= \bm{w}^{\top} \bm{x}/\sqrt{D}$, where the latent task weight $\bm{w} \sim \mathcal{N}(0, \id_D)$.
To construct the $\mu$-th input matrix $\mathbf{X}^\mu$, we use $N$ samples as prompts using the same $\bm{w}^\mu$ yet different $\bm{x}$ within the input sequence. An additional sample $\bm{\tilde{x}}^{\mu}$ is regarded as the query whose true label $\tilde{y}$ is masked yet to be predicted by the transformer. The structure of each input matrix $\mathbf{X}^{\mu}$ is thus represented as
\begin{equation}\label{inputX}
    \mathbf{X}^{\mu}
    =
    \begin{bmatrix}
        \bm{x}_{1}^{\mu} & \bm{x}_{2}^{\mu} & \cdots & \bm{x}_{N}^{\mu} & \bm{\tilde{x}}^{\mu} \\
        y_1^\mu & y_2^\mu & \cdots & y_N^\mu & 0 
    \end{bmatrix} 
    \in \mathbb{R}^{(D+1)\times(N+1)} .
\end{equation}
Due to this form of the input matrix, the number of trainable elements in $\mathbf{W}$ becomes $(D+1)^2$. The last element of $\mathbf{Y}$ corresponds to the predicted label of the query $\bm{\tilde{x}}^{\mu}$, i.e., $\hat{y}^\mu = \mathbf{Y}_{D+1,N+1}^\mu$. The goal of ICL is to use the prompt to form a prediction for the query, and the true function governing the linear relationship for the testing prompt is hidden during pre-training because each $\mu$ is generated by an independently drawn $\bm{w}$ during both training and test phases. \blue{This is in stark contrast to standard supervised learning in perceptron networks that learn a single $\boldsymbol{w}$ (in both training and testing) from examples~\cite{Gyo-1990}}. We consider an ensemble of $P$ sequences, and $P$ is thus called the task diversity. This setting is a bit different from that in recent works~\cite{Ganguli-2024,CP-2024}. 

The pre-training is carried out by minimizing the mean squared error function, and the total training loss is given by
\begin{equation}\label{loss}
    \mathcal{L} = \frac{1}{2P} \sum_{\mu} \left( \tilde{y}^\mu - \hat{y}^\mu \right)^2 + \frac{\lambda}{2} \| \mathbf{W} \|^2 ,
\end{equation}
where $\lambda$ controls the weight-decay strength. The generalization error on unseen tasks is written as \blue{ $\epsilon_{g} = \mathbb{E}_{\bm{\tilde{x}},\bm{w},\mathbf{W}|\mathcal{D},\mathcal{D}} ( \tilde{y} - \hat{y} )^2$, where the ensemble average over all disorders including the pre-training tasks $\mathcal{D}$ is considered}.

\textit{Spin-glass model mapping.}---
Equation~\eqref{loss} can be treated as a Hamiltonian in statistical physics. The linear attention structure makes the spin-model mapping possible.
This proceeds as follows. The prediction to the $\mu$-th input matrix can be recast as $\hat{y}^{\mu} = N^{-1} \sum_{m,n} \mathbf{C}_{D+1,m}^{\mu} \mathbf{W}_{m,n} \mathbf{X}_{n,N+1}^{\mu}$, where $\mathbf{C}^{\mu} \equiv \mathbf{X}^{\mu} \mathbf{X}^{\mu\top}$. Then we define an index mapping $\Gamma:(m,n) \to i$ to flatten a matrix into a vector. \blue{Therefore, one can write $\sigma_i = \Gamma \mathbf{W}_{m,n}$, and $s_i^{\mu} = \Gamma \mathbf{S}_{m,n}^\mu$, where $\mathbf{S}^\mu_{m,n}=N^{-1}\mathbf{C}_{D+1,m}^{\mu} \mathbf{X}_{n,N+1}^{\mu}$}, and $i = (D+1)(m-1)+n$, and finally the prediction as $\hat{y}^{\mu} =\sum_{i} s_i^{\mu} \sigma_i$. Consequently, the mean-squared error for each input matrix reads
\begin{equation}
    \ell^{\mu} = \frac{1}{2} \sum_{i,j} s_i^{\mu} s_j^{\mu} \sigma_i \sigma_j - \tilde{y}^{\mu} \sum_{i} s_i^{\mu} \sigma_i,
\end{equation}
where we omit the constant term $(\tilde{y}^{\mu})^2/2$. 

Upon defining $J_{ij}^{\mu} \equiv - s_i^{\mu} s_j^{\mu}$, $h_i^{\mu} \equiv \tilde{y}^{\mu} s_i^{\mu}$, and $\lambda_i^\mu \equiv \lambda - J_{ii}^\mu$,
 one can rewrite the total loss $\mathcal{L} = (1 / P) \sum_{\mu} \ell^\mu + \frac{1}{2} \lambda \Vert \mathbf{W} \Vert^2$ as 
\begin{equation} \label{eq:Total-Loss}
  \mathcal{L} = - \frac{1}{2P} \sum_{\mu, i,j} J_{ij}^{\mu} \sigma_i \sigma_j - \frac{1}{P} \sum_{\mu, i} h_i^{\mu} \sigma_i + \frac{1}{2} \lambda \sum_{i} \sigma_i^2 .
\end{equation}
By moving the elements for $i=j$ in the first term to the regularization term, and defining an anisotropic regularization coefficient $\lambda_i^\mu \equiv \lambda-J_ {ii}^\mu$, we can formally define the effective interaction $J_{ij} \equiv (1/P) \sum_{\mu} J_{ij}^\mu$, the external field $h_{i} \equiv (1/P) \sum_{\mu} h_{i}^\mu$ and the regularization factor $\lambda_i \equiv (1/P) \sum_{\mu} \lambda_i^\mu$. Finally, we obtain a spin glass model with the following Hamiltonian
\begin{equation} \label{eq:Hamiltonian-P}
  \mathcal{H}(\bm{\sigma}) = - \sum_{i<j} J_{ij} \sigma_i \sigma_j - \sum_{i} h_i \sigma_i + \frac{1}{2} \sum_{i} \lambda_i \sigma_{i}^2,
\end{equation}
where the total number of spins is given by $(D+1)^2$. \blue{According to Eq.~\eqref{loss}, this is precisely the Hamiltonian for a high-dimensional ridge regression interpretation of ICL}.

\begin{figure}
    \centering
\includegraphics[width=0.5\textwidth]{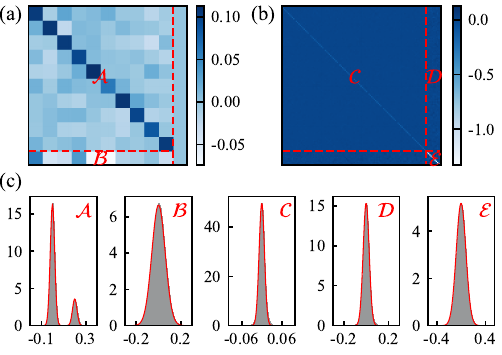}
        \caption{Statistical properties of the interaction matrix and the external field. 
  (a) The $\mathbf{H}$ matrix corresponding to the external field $\boldsymbol{h}$ has three blocks with different properties:  $\mathcal{A} \in \mathbb{R}^{D\times D}$,  $\mathcal{B} \in \mathbb{R}^{D}$, and the all-zero vector $\boldsymbol{0} \in \mathbb{R}^{D+1}$ (the rightmost column). 
  (b) The symmetric $\mathbf{J}$ matrix also has three different blocks: $\mathcal{C} \in \mathbb{R}^{D(D+1)\times D(D+1)}$,  $\mathcal{D} \in \mathbb{R}^{D(D+1)\times (D+1)}$ and $\mathcal{E} \in \mathbb{R}^{(D+1) \times (D+1)}$.
 \blue{(c) The statistics of different blocks in $\mathbf{H}$ and $\mathbf{J}$ matrices. 
  The red line shows a rough Gaussian fitting.
  For some $(i,j)$, $J_{ij}=0$, but this ratio is $(2D+1)/(D+1)^2$ since the last column of $\mathbf{S}$ is an all-zero vector. We count the non-zero values for the distribution density. 
  All results in (c) are plotted based on a collection of $100\,000$ ensembles of $P=1\,000$, $D=10$, and $N=100$.}
    } 
        \label{fig1}
    \end{figure}

The disorder in the pre-training dataset including the diversity in the latent task vectors $\bm{w}$ is now encoded into the interactions between spins, and random fields the spins feel. In fact, this is a densely connected spin glass model, while the coupling and field statistics do not have an analytic expression~\cite{Product-1970}, as they bear a fat tail (Fig.~\ref{fig1}). This reminds us of the two-body spherical spin model studied in spin glass theory~\cite{Sphere-1992,Ricci-2020}, but the current glass model of ICL seems much more complex than the spherical model. By construction, this model reflects the nature of associative memory~\cite{Huang-2022}, yet the spin variable is now the underlying parameter of the transformer.
 To conclude, we derive a spin glass model of ICL, opening a physically appealing route towards the mechanistic interpretation of ICL and even more complex transformers (see extra experiments in~\cite{SM}).

\blue{The specific form of $J_{ij}$ and $h_i$ rely on $\mathbf{S}_{m,n}$ defined as $\mathbf{S}_{m,n} = N^{-1}\mathbf{C}_{D+1,m} \mathbf{X}_{n,N+1}$}. The index $\mu$ is omitted here. The matrix $\mathbf{S}$ is divided into three blocks: the last column is an all-zero vector [due to the masked label in Eq.~\eqref{inputX}], while the other two blocks are labeled as $\mathcal{A}$ ($m<D+1, n\neq D+1$) and $\mathcal{B}$ ($m=D+1, n\neq D+1$).  As $h_i=\frac{1}{P}\sum_{\mu}\tilde{y}^\mu s_i^\mu$,  the field matrix has the same block structure with $\mathbf{S}$ [Fig.~\ref{fig1} (a)]. In addition the interaction matrix $\mathbf{J}$, generated by the outer product of the flattened $\mathbf{S}$ with itself, has three main blocks, labeled as $\mathcal{C}$, $\mathcal{D}$, and $\mathcal{E}$ respectively in Fig.~\ref{fig1} (b). 

In contrast to traditional spin glass models~\cite{Sphere-1992}, $P(\mathbf{J})$ [or $P(\boldsymbol{h})$] does not have an analytic form~\cite{Product-1970}, as the coupling or field can be expressed as a complex function of a sum of products of two i.i.d. standard Gaussian random variables (see details in~\cite{SM}).  Therefore, we provide the numerical estimation of the coupling distribution, which all bear a fat tail \blue{not captured by a Gaussian fitting [Fig.~\ref{fig1} (c)]}. We thus define a new type of spin glass model corresponding to ICL, or a metaphor of transformer in large language models.

\textit{Statistical mechanics analysis.}---
The probability of each weight configuration is given by the Gibbs-Boltzmann distribution $P(\bm{\sigma}) = e^{-\beta \mathcal{H}(\bm{\sigma})} / Z$, where $Z$ is the partition function, and $\beta$ is an inverse temperature tuning the energy level. Because the statistics of couplings and fields have no analytic form, one has to use the cavity method widely used in spin glass theory~\cite{Mezard-1987}. The cavity method is also known as the belief propagation algorithm, working by iteratively solving a closed equation of cavity quantity, i.e., a spin is virtually removed~\cite{Huang-2022}. The cavity method can thus be used on single instances of ICL.  Different weight components are likely strongly correlated, but the cavity marginal $\eta_{i\to j}(\sigma_i)$ becomes conditionally independent in the absence of spin $j$, which facilitates our derivation of the following self-consistent iteration (namely mean-field equation, see~\cite{SM} for more details): 
 \begin{equation} \label{CEq}
 \begin{aligned}
    \eta_{i\to j} (\sigma_i) =& \frac{1}{z_{i \to j}} e^{\beta h_i \sigma_i - \frac{1}{2} \beta \lambda_i \sigma_i^2}  \\
    & \times \prod_{k \neq i,j} \left[\int \mathrm{d} \sigma_k ~ \eta_{k\to i}(\sigma_{k}) ~ e^{\beta J_{ik} \sigma_i \sigma_k}\right] ,
\end{aligned}
\end{equation}
where $z_{i\to j}$ is a normalization constant, and $\eta_{i\to j}$ is defined as the cavity probability of spin $\sigma_i$ in the absence of the interaction between spins $i$ and $j$. After the iteration reaches a fixed point, the marginal probability $\eta_{i}(\sigma_i)$ of each spin can be calculated by
\begin{equation}
    \eta_{i} (\sigma_i) = \frac{1}{z_i} \mathrm{e}^{\beta  h_i \sigma_i - \frac{1}{2} \beta \lambda_i \sigma_i^2} \prod_{j \neq i} \int \mathrm{d}\sigma_j ~ \eta_{j\to i}(\sigma_{j}) ~ \mathrm{e}^{\beta J_{ij} \sigma_i \sigma_j} .
\end{equation}

Because of the continuous nature of spin and weak but dense interactions among spins,  we can further simplify the mean-field equation [Eq.~\eqref{CEq}], and derive the approximate message passing (AMP) by assuming $\eta_{i}(\sigma_i) \sim \mathcal{N}(m_i, v_i)$, where $(m_i, v_i)$ is the fixed point of the following iterative equation:
\begin{subequations}
\begin{align}
     m_{i}&= \frac{\beta h_i + \beta \sum_{j \neq i} J_{ij} m_{j}}{\beta \lambda_i - \beta^2 \sum_{j \neq i} J^2_{ij} v_{j}} , \label{eq:AMP_m} \\
     v_{i} &= \frac{1}{\beta \lambda_i - \beta^2 \sum_{j \neq i} J^2_{ij} v_{j}}, \label{eq:AMP_v}
\end{align}
\end{subequations}
which is also rooted in the Thouless–Anderson–Palmer equation in glass physics~\cite{TAP-1977,Huang-2022}. Technical details of deriving the AMP and the self-consistent justification of the Gaussian approximation are given in the appendix (an expanded one is given in~\cite{SM}).

\textit{Results.}---
\blue{In this paper, we consider a non-asymptotic setting: $D$ is kept finite and independent of $P$ and $N$. }
\begin{figure}
	\centering
	\includegraphics[width=0.5\textwidth]{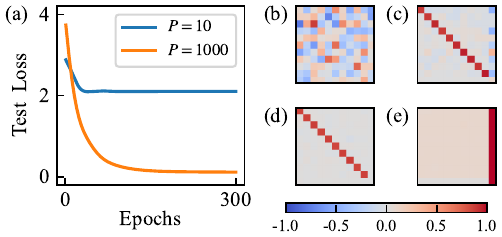}
	\caption{
  Test error and the optimal weight matrix of ICL. 
    (a) The test error of the linear attention trained by the stochastic gradient descent (SGD) method for different task diversities.
    (b) The weight matrix for $P=10$ at the end of training.
    (c) The weight matrix for $P=1\,000$ at the end of training.
    (d) The weight matrix retrieved from the solution $\{m_i\}$ of AMP.
    (e) The variance matrix retrieved from the solution $\{v_i\}$ of AMP. 
    Parameters: $(D, N)=(10, 100)$, $\lambda=0$ for SGD, $P=5\,000$, $(\lambda,\beta)=(0.01,100)$ for AMP. 
	}\label{fig2}
\end{figure}

\begin{figure}
	\centering
	\includegraphics[width=0.5\textwidth]{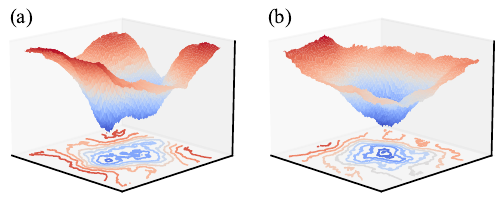}
	\caption{The energy landscape of the spin glass model for $P=1\,000$ (a) and $P=10$ (b). To draw this landscape, we randomly sample $1\,000$ points from the whole weight space when $D=10$, $N=50$, $\lambda=0.01$, and calculate their energies by Eq.~\eqref{eq:Hamiltonian-P}. Then we reduce the weight space with $(D+1)^2$ dimension to the two-dimensional plane by t-distributed stochastic neighbor embedding (t-SNE)~\cite{Hinton-2008} and plot the energy surface by radial basis function interpolation with the linear kernel. The color deep blue indicates the lowest energies. 
	}\label{fig3}
\end{figure}

To see whether our spin glass model captures the correct structure of the weight matrix in the simple transformer, we 
first divide the weight matrix $\mathbf{W}$ into blocks in the same way as we do for the input matrix, i.e.,
\begin{equation}
\mathbf{W} = 
\begin{bmatrix}
    \mathbf{W}_{11} & \mathbf{W}_{12} \\
    \mathbf{W}_{21} & \mathbf{W}_{22}
\end{bmatrix} 
\end{equation}
where $\mathbf{W}_{11} \in \mathbb{R}^{D\times D}$, $\mathbf{W}_{12} \in \mathbb{R}^{D\times 1}$, $\mathbf{W}_{21} \in \mathbb{R}^{1 \times D}$, and $\mathbf{W}_{22} \in \mathbb{R}$. 
\blue{With a large $N$, the actual prediction of the transformer to the test query $\tilde{\boldsymbol{x}}$ can be written as
\begin{equation}
    \hat{y} =\frac{1}{\sqrt{D}} \boldsymbol{w}^\top \left( \mathbf{W}_{11} +\frac{1}{\sqrt{D}} \boldsymbol{w} \mathbf{W}_{21} \right) \tilde{\boldsymbol{x}}.
\end{equation}
}
 Hence, \blue{to get rid of post-training for a new task vector $\boldsymbol{w}$}, the weight matrix of a well-pretrained transformer must satisfy
\blue{
\begin{equation}
    \mathbf{W}_{11} + \frac{1}{\sqrt{D}}\boldsymbol{w} \mathbf{W}_{21} = \id_D.
\end{equation}
}
To derive the above equations, we have used $2\times 2$ block matrix form of $\mathbf{X}$, the task \blue{$y= \bm{w}^{\top} \bm{x} / \sqrt{D}$, and $\mathbf{X}_0 \mathbf{X}_0^\top \approx N \id_D$ for large $N$ but finite $D$, where $\mathbf{X}_0 = [ \bm{x}_{1}, \bm{x}_{2}, \cdots, \bm{x}_{N} ]$. (see technical details in SM~\cite{SM}). In the case of $P>1$, the weights have a unique optimal solution $\mathbf{W}_{11} = \id_D$ and $\mathbf{W}_{21} = 0$. However, the minimum of the loss is not guaranteed to be unique if the Hamiltonian has zero modes, bearing similarity with Boltzmann machine learning~\cite{Remi-2012}. The number of zero modes decreases with $P$ (see SM~\cite{SM}), supported by the following results (Fig.~\ref{fig2} and Fig.~\ref{fig3}) as well. }

In the standard stochastic gradient descent (SGD) training process minimizing Eq.~\eqref{loss}, a large value of $P$ is needed to make the weight matrix converge to the unique solution. In Fig.~\ref{fig2}, we show the learning curves and the weight matrix after the training when the amount of training data $P=10$ and $P=1\,000$ respectively. By iterating the AMP equations [Eq.~\eqref{eq:AMP_m} and Eq.~\eqref{eq:AMP_v}], we get a fixed point of $\{m_{i}\}$ and $\{v_{i}\}$, transformed back into the matrix form by inverting $\Gamma$. We find that the $\mathbf{m}$ matrix exhibits the same property as the weight matrix that is well-trained by the SGD. Since the last column of $\mathbf{W}$ is initialized as $\mathcal{N}(0,1)$ and does not participate in the training, the solution of AMP retains the structure of $m=0$ and $v=1$. This result shows that our spin glass model captures the properties of practical SGD training~\cite{Zhang-2024,Chen-2024,Conv-2023}. \blue{However, a separate training of query and key matrices makes the learning dynamics different, while yielding the same performance. Even if using a rough Gaussian fitting [Fig.~\ref{fig1} (c)] to the model parameters (the bulk is captured but not the tail), we get the diagonal structure of the weight matrix (see details in SM~\cite{SM}). }

\begin{figure}
	\centering
	\includegraphics[width=0.5\textwidth]{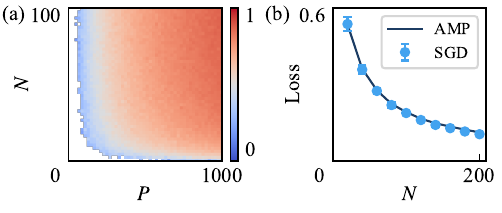}
	\caption{Results obtained from running the AMP algorithm.
    (a) The heat map of the contrast $\mathscr{C}$ with $D=20$, $\lambda=1$, and $\beta=100$. In the left-bottom corner, AMP does not converge.
    (b) The test error decreases with the context (prompt) length $N$, with $D=20$, $\lambda=0.01$, $P=10\,000$, and $\beta=100$. 
    All the results of AMP and SGD are averaged over $50$ trials.
        }\label{fig4}
\end{figure}

In Fig.~\ref{fig3}, we show the energy landscape of our spin glass model. When the task diversity $P$ is large enough,
 there is only one global minimum in the energy landscape, and the learning can easily reach the lowest energy. When $P$ is small, 
 there emerge multiple local minima in the energy landscape, and the learning gets easily trapped by metastable states, which prevents the transformer from accurate in-context inference.
 \blue{We also support this observation by an analysis of the spectral density of the Hamiltonian's Hessian matrix which displays a number of zero modes in SM~\cite{SM}}.

To get the phase diagram for single instance pre-training of the transformer, \blue{ we define the contrast ratio as the cosine similarity between the $\mathbf{m}$ matrix and the identity matrix: $\mathscr{C} = \left[ \langle \mathbf{m}, \id_D \rangle / (\Vert \mathbf{m} \Vert^2 \Vert \id_D \Vert^2 )\right]$, where $\langle \mathbf{A}, \mathbf{B} \rangle = \sum_{ij} \mathbf{A}_{ij} \mathbf{B}_{ij}$, $\Vert \mathbf{A} \Vert^2$ is the matrix Frobenius norm, and $[\cdots]$ denotes the average over different realizations. $\mathscr{C}$ measures whether the model is well trained according to the transformed $\mathbf{m}$ matrix.} $\mathscr{C} = 1$ means that the model converges to the unique solution, while $\mathscr{C} = 0$ indicates that the model does not learn the features at all. We show a heat map of the contrast ratio with \blue{the task diversity $P$ and the prompt length $N$} in Fig.~\ref{fig4} (a). The heat map suggests that when the task diversity increases, a smooth transition to perfect generalization occurs, while keeping a large value of task diversity, increasing the prompt length further lowers the generalization error, which is consistent with recent empirical works~\cite{Ganguli-2024} and theoretical works~\cite{CP-2024} based on random matrix theory (despite a slightly different setting). In addition, the AMP result coincides perfectly with the SGD (Fig.~\ref{fig4} (b), \blue{see also the case of non-i.i.d. data in SM~\cite{SM}}), which verifies once again that our spin-glass model of ICL is able to predict an unseen embedded function in the test prompts, which is determined by the ground states of $\mathcal{H}(\boldsymbol{\sigma})$. We finally remark that our theory carries over to
more complex situations of ICL (see further experiments in~\cite{SM}).

\textit{Conclusion.}---
A fundamental question in large language models is what contributes to the emergence ability of ICL, i.e., why simple next-token prediction-based pre-training leads to in-context learning of previously unseen tasks, especially without further tuning the model parameters. Here, we turn the ICL into a spin glass model and verify the equivalence between the standard SGD training and our statistical mechanic inference. We observe the fat tail distribution of coupling that determines how the model parameters of the transformer interact with each other. The transformer parameters are akin to an ensemble of real-valued spins in physics whose ground state suggests that the model can infer an unknown function from the shown test prompts after a pre-training of input sequences of sufficient task diversity. The phase diagram for single instance learning is also derived by our method, suggesting a continuous ICL transition.

The spin-glass model mapping of the linear transformer establishes a toy model of understanding emergent abilities such as ICL of large language models. The ground state determines the intelligence boundary, while the task diversity guarantees the accessibility of the ground state. Without a clear understanding of this toy model, it is hard to imagine what is really computed inside the black box of a general transformer in more complex tasks. Future exciting directions include explaining the chain-of-thought prompting~\cite{Wei-2022}, i.e., decomposition of a complex task into intermediate steps, and more challenging case of hallucination~\cite{Nature-2024}, i.e., the model could not distinguish the generated outputs from factual knowledge, or it could not understand what they generate~\cite{Huang-2024}. We speculate that this hallucination may be intimately related to the solution space of the spin glass model given a fixed complexity of training dataset, e.g., spurious states in a standard associative memory model, as implied by Eq.~\eqref{eq:Hamiltonian-P}. These open questions are expected to be addressed in the near future by considering this intriguing physics link, thereby enhancing the robustness and trustworthiness of AI systems.

\section*{Acknowledgments}
 This research was supported by the National Natural Science Foundation of China for
Grant number 12475045 and 12122515, and Guangdong Provincial Key Laboratory of Magnetoelectric Physics and Devices (No. 2022B1212010008), and Guangdong Basic and Applied Basic Research Foundation (Grant No. 2023B1515040023).

\onecolumngrid
 \setcounter{figure}{0}    
\renewcommand{\thefigure}{S\arabic{figure}}
\renewcommand\theequation{S\arabic{equation}}
\setcounter{equation}{0}    

\setcounter{figure}{0}    
\renewcommand{\thefigure}{S\arabic{figure}}
\renewcommand\theequation{S\arabic{equation}}
\setcounter{equation}{0}  
\appendix
\section{Comparison of trainings with merged weight and separated weight matrices}

\begin{figure}[b]
  \centering
  \includegraphics{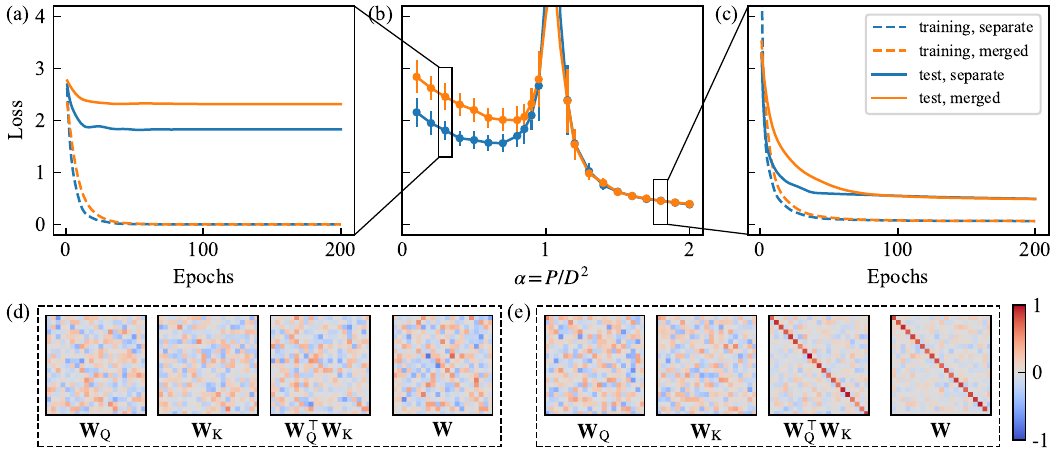}
  \caption{
    Losses and weights obtained by SGD training for the merged weight and separated weight schemes.
    (a) The training and test losses for the data-poor case ($\alpha = 0.3$). Note that $D$ is fixed (non-asymptotic limit).
    (b) The test errors of the two schemes show the double descent phenomenon as the diversity increases.
    (c) The training and test losses for the data-rich case ($\alpha = 1.8$).
    (d) The trained weight matrices of the two schemes, corresponding to the end of the dynamics in (a).
    (e) The trained weight matrices corresponding to the end of the dynamics in (c).
    Parameters: $D = 20$, $N = 100$, $\lambda = 0$ for all subfigures. The results in (b) are averaged over $50$ trials.
  } 
\label{fig-S1}
\end{figure}

In this section, we discuss the difference in training between the merged weight matrix $\mathbf{W}$ used in the main text and the separated weight matrix $\mathbf{W}_{\rm Q}$ and $\mathbf{W}_{\rm K}$ in the linear attention model. To be more precise, we consider the linear attention with the merged weight as
\begin{equation}
  \mathbf{Y} = \frac{1}{N} \mathbf{XX}^\top \mathbf{WX} ,
\end{equation}
and the linear attention with the separated weight as
\begin{equation}
  \mathbf{Y} = \frac{1}{N} \mathbf{XX}^\top \mathbf{W}_{\rm Q}^\top \mathbf{W}_{\rm K} \mathbf{X} .
\end{equation}

The generalization error of the two schemes shows the expected double descent phenomenon, in the ridgeless limit $\lambda \to 0$, as shown in Fig.~\ref{fig-S1}\,(b). In the data-poor case, i.e., the data density $\alpha \equiv P / D^2 < 1$, the scheme with separated weights ultimately exhibits lower generalization errors, while the dynamics of the training process are similar for both schemes [see Fig.~\ref{fig-S1}\,(a)]. In the data-rich case, the performances of the two schemes are highly similar, except that the scheme with separated weights converges a bit faster on the test tasks [see Fig.~\ref{fig-S1}\,(c)]. Note also that we fix $D=20$ and thus this is not an asymptotic limit, being the same setting we focus on in the main text.

For the trained weight matrices, we find that in the data-poor case, neither scheme captures the correct weight structure [see Fig.~\ref{fig-S1}\,(d)]; while in the data-rich case, the two training schemes lead to a similar diagonal structure, which is the optimal solution of the problem [see Fig.~\ref{fig-S1}\,(e)]. This justifies the use of the scheme of the merged matrix in the main text.

\section{Statistical properties of the interactions and external fields}

In this section, we provide detailed statistical analyses of the interaction matrix $\mathbf{J}$ and the external field $\mathbf{h}$ and also test the robustness of our
spin-glass mapping against the non-i.i.d. data setting.

\subsection{Explicit formulation of the model parameters}

We present here the explicit formulation of the elements in each block in Fig.~1\,(c) in the main text.
The input matrix $\mathbf{X}$ can be written into a $2\times2$ block matrix:
\begin{equation}
    \mathbf{X} = 
    \left[\begin{array}{cc}
      \mathbf{X}_0 & \boldsymbol{\tilde{x}} \\
      \boldsymbol{y}_0^\top  & 0 
    \end{array}\right] ,
\end{equation}
where $\mathbf{X}_0 = [ \bm{x}_{1}, \bm{x}_{2}, \cdots, \bm{x}_{N} ] \in \mathbb{R}^{D\times N}$ denotes the data used for the prompts, $\bm{y}_0 = \mathbf{X}_0^\top \bm{w}/\sqrt{D}$ is the corresponding label vector, and $\boldsymbol{\tilde{x}}$ is the testing prompt whose label needs to be predicted by the transformer. Thus, we can calculate
\begin{equation} \label{eq:XXT}
    \mathbf{XX^\top} = 
    \left[\begin{array}{cc}
        \mathbf{X}_0 & \boldsymbol{\tilde{x}} \\
        \boldsymbol{y}_0^\top  & 0 
    \end{array}\right]
    \left[\begin{array}{cc}
        \mathbf{X}_0^\top & \boldsymbol{y}_0 \\
        \boldsymbol{\tilde{x}}^\top  & 0 
    \end{array}\right]
    = 
    \left[\begin{array}{cc}
        \mathbf{X}_0 \mathbf{X}_0^\top + \boldsymbol{\tilde{x}} \boldsymbol{\tilde{x}}^\top & \mathbf{X}_0 \boldsymbol{y}_0  \\
        \boldsymbol{y}_0^\top \mathbf{X}_0^\top  & \boldsymbol{y}_0^\top \boldsymbol{y}_0
    \end{array}\right] .
\end{equation}
Recalling the definition of $\mathbf{S}^\mu_{mn} = N^{-1} \sum_{i} \mathbf{X}_{D+1,i}^{\mu} \mathbf{X}_{m,i}^{\mu} \mathbf{X}_{n,N+1}^{\mu}$, we can rewrite the $\mathbf{S}$ matrix as
\begin{equation}
  \mathbf{S} = \frac{1}{N} 
  \left[\begin{array}{c}
  \mathbf{X}_0 \boldsymbol{y}_0 \\
  \boldsymbol{y}_0^\top \boldsymbol{y}_0
  \end{array}\right]
  \otimes
  \left[\begin{array}{c}
  \boldsymbol{\tilde{x}} \\
  0
  \end{array}\right]
  =
  \frac{1}{N}
  \left[\begin{array}{cc}
  \mathbf{X}_0 \boldsymbol{y}_0 \boldsymbol{\tilde{x}}^\top & 0 \\ 
  \boldsymbol{y}_0^\top \boldsymbol{y}_0 \, \boldsymbol{\tilde{x}}^\top & 0
  \end{array}\right] ,
\end{equation}
where $  \otimes$ denotes an outer product of two vectors.
Since $h_i=\frac{1}{P}\sum_\mu\tilde{y}^\mu s_i^\mu$, where $s_i^\mu$ is a vectorization of $\mathbf{S}^\mu$,
 we can write expressions for the elements in blocks $\mathcal{A}$ and $\mathcal{B}$, which we denote as $A$ and $B$ respectively:
\begin{subequations} \label{eq:AB}
\begin{align}
  & A_{ij} = \frac{1}{PDN} \sum_{\mu,k} w_k \tilde{x}_k^{\mu} \sum_{m} x_{im}^{\mu} \sum_{n} w_{n} x_{nm}^{\mu} \tilde{x}_{j}^{\mu} , \\
  & B_{i} = \frac{1}{PDN} \sum_{\mu,k} w_k \tilde{x}_k^{\mu} \sum_{m} \Big( \sum_{n} w_{n} x_{nm}^{\mu} \Big)^2 \tilde{x}_{i}^{\mu} .
\end{align}
\end{subequations}
All the variables $w, x, \tilde{x}$ used above are i.i.d. standard Gaussian random variables. We denote the distribution of elements $\{A_{ij}\}$ in block $\mathcal{A}$ as $P(A)$, and those in block $\mathcal{B}$ as $P(B)$, which are both complex combinations of the sum and product of two i.i.d. Gaussian random variables. It is computationally challenging to conclude an analytic form for the distribution of random variables specified in Eq~\eqref{eq:AB}. In this way, we can also write the elements in other blocks as: $C = A_1 A_2$, $D = A_1 B_1$ and $E = B_1 B_2$, where $A_1, A_2 \sim P(A)$, $B_1, B_2 \sim P(B)$.

\begin{figure}[h]
  \centering
  \includegraphics{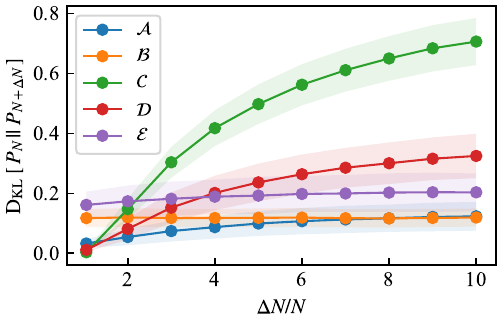}
  \caption{The KL divergence $\text{D}_{\text{KL}} [P_{N} \| P_{N+\Delta N}]$ between the original distribution of the elements in blocks shown in Fig.~1 in the main text and that with the prompt length $N$ prolonged by $\Delta N$. The results are averaged over $1\,000$ examples with $(P,D,N)=(1\,000,20,20)$.} 
  \label{fig-S2}
\end{figure}

A surprising observation is that the distributions of blocks $\mathcal{A}$, $\mathcal{C}$ and $\mathcal{D}$ seem sensitive to the change of the prompt length $N$ (see Fig.~\ref{fig-S2}, but a weaker effect for $\mathcal{B}$ and $\mathcal{E}$). However, once $P$ is sufficiently large, the correct weight solution will be learned, and thus the transformer achieves a nearly perfect in-context inference, which captures the essence of ICL explained in the main text. This demonstrates that even if the parameters of the Hamiltonian vary, the ground state remains the same for the ICL.

\subsection{Running AMP given a fitted distribution of model parameters}
Although an analytic form of the model parameters is unknown, we fit the experimental data by a Gaussian (or mixture) distribution (not accurate due to the tail, but the Cauchy distribution does not capture the bulk very well). Then we sample the couplings and fields from the fitted distribution and run the AMP equation to solve the spin model Eq.~(5) in the main text. This is the standard setting of a spin glass problem as the model parameters are sampled from an analytic distribution. However, in the ICL, the model parameters are directly constructed from the supplied dataset, reflecting the intrinsic structural disorder in the training tasks. We address how different these two scenarios are. 

\begin{figure}[h]
  \centering
  \includegraphics{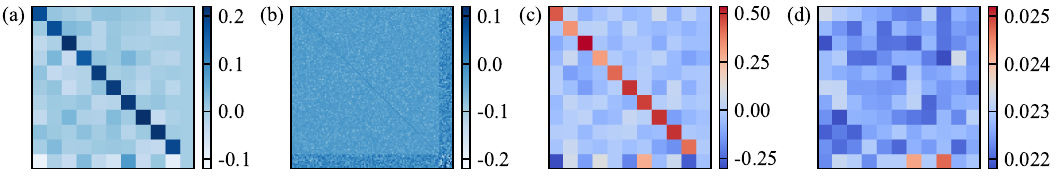}
  \caption{
    (a) The $\mathbf{H}$ matrix generated by resampling from the fitted Gaussian distribution (see Fig.~1 (c) in the main text).
    (b) The $\mathbf{J}$ matrix generated by resampling from the fitted Gaussian distribution  (see Fig.~1 (c) in the main text).
    (c) The mean $\mathbf{m}$ of the weight matrix obtained by solving the AMP equation.
    (d) The variance $\mathbf{v}$ of the weight matrix obtained by solving the AMP equation.
    Parameters: $P=1\,000$, $D=10$, $N=100$, and $\lambda = 1$.
  } 
  \label{fig-S3}
\end{figure}

The fitting distributions are shown by the red line in Fig.~1\,(c) in the main text, in the case of $P=1\,000$, $D=10$, $N=100$. Then we resample randomly from the fitting distribution to generate $\mathbf{J}$ and $\boldsymbol{h}$, as shown in Fig.~\ref{fig-S3}\,(a)(b). We then run the AMP equation on these parameters, and the results are shown in Fig.~\ref{fig-S3}\,(c)(d), which indicates that AMP still captures the diagonal structure of the optimal solution, although there are some numerical deviations in the diagonal elements. Even if we discard some information in the real distribution, we recover the basic feature of the spin glass model for ICL.

\subsection{Robustness of the spin model mapping to non-i.i.d. data}

\begin{figure}
    \centering
    \subfigure[]{
        \includegraphics[width=0.5\textwidth]{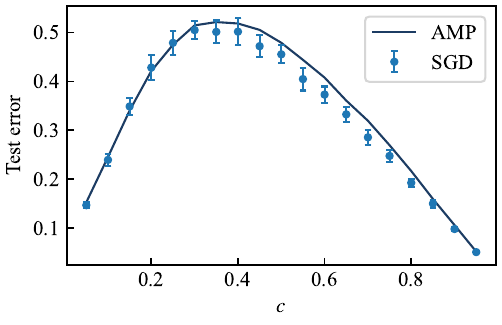}
   }
   \subfigure[]{
        \includegraphics[width=0.7\textwidth]{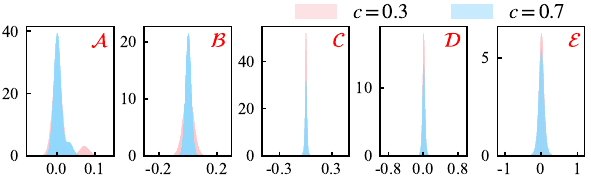}
       }
  \caption{Training with correlated data. (a)
    Test errors of AMP and SGD vary with correlation degree $c$. $P=10\,000$, $D=20$, $N=200$, $\lambda = 0.01$, and $\beta=100$.
     All results are averaged over $20$ trials. (b) Distributions of couplings and fields as in Fig.~1 (c) in the main text.
  } 
  \label{fig-S4}
\end{figure}

We test whether our spin model is still valid on a correlated data set. The columns in the data matrix $\mathbf{X}_0$ are no longer i.i.d., but correlated, which models the intrinsic relationship between tokens in real language tasks. Specifically, the correlated data samples are constructed by the following rule~\cite{JF-1989}:
\begin{equation}\label{corr}
  \left< x_{i}^\mu x_{k}^\nu \right> = \delta_{ik} \left[ c + (1-c) \delta_{\mu\nu} \right],
\end{equation}
where $\mu$ indicates different columns in the input $\mathbf{X}$, and $i$ indicates different rows. This correlation design means that elements in different rows are independent of each other for each column, and the covariance of elements in the same row is regulated by $c$. Keeping the other settings unchanged, we plot the performances of SGD and AMP in Fig.~\ref{fig-S4}. Even though the generalization error is not simply monotonic with $c$, the results of AMP are still able to match those of SGD. The distributions of couplings and fields for the spherical spin model mapping look similar to the i.i.d. scenario. 

\section{Derivation of cavity method and approximate message passing equation}

The spin glass model of the transformer we consider in the main text reads
\begin{equation} \label{eq:Hamiltonian}
    H = - \sum_{i<j} J_{ij} \sigma_i \sigma_j - \sum_i h_i \sigma_i - \frac{\lambda}{2} \sum_i \sigma_i^2 ,
\end{equation}
and the Boltzmann distribution of weight configuration can be written as
\begin{equation} \label{eq:Boltzmann}
    P(\bm{\sigma}) = \frac{1}{Z} e^{- \beta H(\bm{\sigma})} = \frac{1}{Z} \prod_{i} e^{\beta h_i \sigma_i - \frac{\beta \lambda}{2} \sigma_i^2} \prod_{i<j} e^{\beta J_{ij} \sigma_i \sigma_j} ,
\end{equation}
where $Z$ serves as the partition function.
Taking each interaction pair as a factor node and each site term as an external field in Eq.~\eqref{eq:Boltzmann}, one can write the following cavity iteration in an explicit form according to the standard format of 
the cavity method (e.g., see the textbook~\cite{Huang-2022}).
\begin{align} 
    & \eta_{i\to ij} (\sigma_i) = \frac{1}{z_{i \to ij}} \mathrm{e}^{\beta  h_i \sigma_i - \frac{\beta \lambda}{2} \sigma_i^2} \prod_{ik \in \partial i \backslash ij} \eta_{ik \to i}(\sigma_{b}) , \label{eq:BP-1} \\
    & \eta_{ij \to i} (\sigma_i) = \frac{1}{z_{ij \to i}} \int \prod_{j\in \partial ij \backslash i} \mathrm{d} \sigma_j ~ \eta_{j\to ij}(\sigma_{j}) ~ \mathrm{e}^{\beta J_{ij} \sigma_i \sigma_j} , \label{eq:BP-2}
\end{align}
which is also called the belief propagation equation. The notation $ij$ denotes an interaction (or factor node in a factor graph representation of the model). Since our model only involves two-body interactions, which means that $\partial ij \backslash i$ contains only one element $j$, the product notation $\prod_{j\in \partial ij \backslash i}$ can be omitted, and
 we can simplify the notation $i\to ij$ by $i\to j$ and $ij\to i$ by $j\to i$. Therefore, we can combine Eq.~\eqref{eq:BP-1} and Eq.~\eqref{eq:BP-2} together into
\begin{equation} \label{eq:BP-3}
    \eta_{i\to j} (\sigma_i) = \frac{1}{z_{i \to j}} e^{\beta h_i \sigma_i - \frac{1}{2} \beta \lambda_i \sigma_i^2} \prod_{k \neq i,j} \left[\int \mathrm{d} \sigma_k ~ \eta_{k\to i}(\sigma_{k}) ~ e^{\beta J_{ik} \sigma_i \sigma_k}\right].
\end{equation}
After the iteration reaches a fixed point, the marginal probability $\eta_{i}(\sigma_i)$ of each spin can be calculated based on the converged cavity marginals as follows
\begin{equation} \label{eq:BP-4}
    \eta_{i} (\sigma_i) = \frac{1}{z_i} \mathrm{e}^{\beta  h_i \sigma_i - \frac{1}{2} \beta \lambda_i \sigma_i^2} \prod_{j \neq i} \int \mathrm{d}\sigma_j ~ \eta_{j\to i}(\sigma_{j}) ~ \mathrm{e}^{\beta J_{ij} \sigma_i \sigma_j} .
\end{equation}

\subsection{Relaxed belief propagation}

The space complexity to run Eq.~\eqref{eq:BP-3} is of $\mathcal{O}(D^4)$, since $i=1,\cdots,(D+1)^2$. To reduce the space complexity, an intuitive approach is to approximate the cavity marginal by a Gaussian distribution $\mathcal{N}(m_{i\to j}, v_{i\to j})$, and then derive the iterative equations for the first two moments. To validate this intuition, we start from Eq.~\eqref{eq:BP-3} and apply the Fourier transform and Taylor expansion.

To proceed, we first define $\xi_{i\to j} = \sum_{k \neq i, j} \beta J_{ik} \sigma_k$, and $G(\xi_{i \to j}) = \mathrm{e}^{\sigma_i \xi_{i \to j}}$, and then Eq.~\eqref{eq:BP-3} can be written as
\begin{subequations} \label{eq:rBP-1}
\begin{align}
    \eta_{i\to j} (\sigma_i) 
    &= \frac{1}{z_{i \to j}} \mathrm{e}^{\beta  h_i \sigma_i - \frac{1}{2} \beta \lambda_i \sigma_i^2} \int \prod_{k\neq i, j} \left[ \mathrm{d} \sigma_k ~ \eta_{k\to i}(\sigma_{k})\right] G(\xi_{i \to j}) \label{eq:rBP-1a} \\
    &= \frac{1}{z_{i \to j}} \mathrm{e}^{\beta  h_i \sigma_i - \frac{1}{2} \beta \lambda_i \sigma_i^2}  \int \prod_{k\neq i, j} \left[ \mathrm{d} \sigma_k ~ \eta_{k\to i}(\sigma_{k})\right] ~ \int \mathrm{d} \hat{\xi}_{i \to j} ~ \hat{G} (\hat{\xi}_{i \to j}) ~ e^{i \xi_{i \to j} \hat{\xi}_{i \to j}} \label{eq:rBP-1b} \\
    &= \frac{1}{z_{i \to j}} \mathrm{e}^{\beta  h_i \sigma_i - \frac{1}{2} \beta \lambda_i \sigma_i^2} \int \mathrm{d} \hat{\xi}_{i \to j} \hat{G} (\hat{\xi}_{i \to j}) \prod_{k\neq i, j} \int \mathrm{d} \sigma_k ~ \eta_{k\to i}(\sigma_{k}) ~ \exp \Big[ i \hat{\xi}_{i \to j} \left( \beta J_{ik} \sigma_k \right) \Big]  \label{eq:rBP-1c}.
\end{align}
\end{subequations}
In Eq.~\eqref{eq:rBP-1b}, we insert the Fourier transform of $G(\xi)$, i.e., $\hat{G}(\hat{\xi}_{i\to j})$, and absorb irrelevant constants into $z_{i\to j}$. Let us calculate the last integral in Eq.~\eqref{eq:rBP-1c}:
\begin{subequations} \label{eq:rBP-2}
\begin{align}
    \mathcal{I} 
    &\equiv \int \mathrm{d}\sigma_k ~ \eta_{k\to i}(\sigma_{k})  ~ \exp \Big[ i \hat\xi \left( \beta J_{ik} \sigma_k \right) \Big] \label{eq:rBP-2a} \\
    &= \int \mathrm{d}\sigma_k ~ \eta_{k\to i}(\sigma_{k}) \left[ 1 +  i \hat\xi \left( \beta J_{ik} \sigma_k \right) - \frac{1}{2} \hat{\xi}^2 \left(\beta J_{ik} \sigma_k \right)^2 \right] \label{eq:rBP-2b} \\
    &= 1 + i \hat\xi \beta J_{ik}  \int \mathrm{d}\sigma_k ~ \eta_{k\to i}(\sigma_{k}) \sigma_k - \frac{1}{2} \hat{\xi}^2 \beta^2 J^2_{ik}  \int \mathrm{d}\sigma_k ~ \eta_{k\to i}(\sigma_{k}) \sigma_k^2 \label{eq:rBP-2c} \\
    &= 1 + i \hat\xi \beta J_{ik}  m_{k \to i} - \frac{1}{2} \hat{\xi}^2 \beta^2 J^2_{ik}  \left( v_{k \to i} + m_{k \to i}^2 \right) \label{eq:rBP-2d} \\
    &\simeq \exp \left[ i \hat\xi \beta J_{ik} m_{k \to i} - \frac{1}{2} \hat{\xi}^2 \beta^2 J^2_{ik} v_{k \to i} \right] \label{eq:rBP-2e} .
\end{align}
\end{subequations}
In Eq.~\eqref{eq:rBP-2b}, we use the Taylor expansion of an exponential function, and in Eq.~\eqref{eq:rBP-2c}, we define the mean $m_{k\to i}$ and the variance $v_{k\to i}$ of the message (the cavity marginal probability) as
\begin{align}
    & m_{k \to i} = \int \mathrm{d} \sigma_{k} ~ \eta_{k \to i}(\sigma_{k}) ~ \sigma_{k} , \\
    & v_{k \to i} = \int \mathrm{d} \sigma_{k} ~ \eta_{k \to i} (\sigma_{k}) ~ \sigma_{k}^2 - m_{k \to i}^2.
\end{align}
In Eq.~\eqref{eq:rBP-2e}, we use the fact that the high order power of $J_{ij}$ are negligible and thus recover the exponential form. Then, we substitute Eq.~\eqref{eq:rBP-2} back to Eq.~\eqref{eq:rBP-1} and obtain
\begin{equation} \label{eq:rBP-3}
    \eta_{i\to j} (\sigma_i) = \frac{1}{z_{i \to j}} \mathrm{e}^{\beta h_i \sigma_i - \frac{1}{2} \beta \lambda_i \sigma_i^2} \int \mathrm{d} \hat{\xi}_{i \to j} \hat{G} (\hat{\xi}_{i \to j}) \prod_{k\neq i, j} \exp \left[ i \hat{\xi}_{i \to j} \beta J_{ik} m_{k \to i} - \frac{1}{2} \hat{\xi}_{i \to j}^2 \beta^2 J^2_{ik} v_{k \to i} \right] .
\end{equation}

Then, we re-express $\hat{G} (\hat{\xi}_{i \to j})$ by $G(\xi)$ and obtain
\begin{subequations} \label{eq:rBP-4}
\begin{align}
    \eta_{i\to j} (\sigma_i) 
    &= \frac{1}{z_{i \to j}} \mathrm{e}^{\beta h_i \sigma_i - \frac{1}{2} \beta \lambda_i \sigma_i^2} \int \mathrm{d} \hat{\xi}_{i \to j} \int \mathrm{d} \xi_{i \to j} ~ G(\xi_{i \to j}) ~ e^{-i \xi_{i \to j} \hat{\xi}_{i \to j}} \prod_{k\neq i, j} \exp \left[ i \hat{\xi}_{i \to j} \beta J_{ik} m_{k \to i} - \frac{1}{2} \hat{\xi}_{i \to j}^2 \beta^2 J^2_{ik} v_{k \to i} \right] \\
    &= \frac{1}{z_{i \to j}} \mathrm{e}^{\beta h_i \sigma_i - \frac{1}{2} \beta \lambda_i \sigma_i^2} \int \mathrm{d} \xi_{i \to j} ~ G(\xi_{i \to j}) \notag \\ 
    &  \times \int \mathrm{d} \hat{\xi}_{i \to j} \exp \bigg[ - \frac{1}{2} \Big( \beta^2 \sum_{k\neq i, j} J^2_{ik} v_{k \to i} \Big) \hat{\xi}_{i \to j}^2 + i \Big( \beta \sum_{k\neq i, j} J_{ik} m_{k \to i} - \xi_{i \to j} \Big) \hat{\xi}_{i \to j} \bigg]  \\ 
    &= \frac{1}{z_{i \to j}} \mathrm{e}^{\beta h_i \sigma_i - \frac{1}{2} \beta \lambda_i \sigma_i^2} \int \mathrm{d} \xi_{i \to j} ~ G(\xi_{i \to j}) \exp \Bigg\{ - \frac{1}{2} \frac{\left( \xi_{i \to j} - \beta \sum_{k\neq i, j} J_{ik} m_{k \to i} \right)^2}{\beta^2 \sum_{k\neq i, j} J^2_{ik} v_{k \to i}} \Bigg\}.
\end{align}
\end{subequations}
We can then define the mean and variance of $\xi_{i\to j}$ as
\begin{align}
    & M_{i \to j} = \beta \sum_{k\neq i, j} J_{ik} m_{k \to i} , \\
    & V_{i \to j} = \beta^2 \sum_{k\neq i, j} J^2_{ik} v_{k \to i} .
\end{align}
Then, the marginal probability $\eta_{i\to j}(\sigma_i)$ can be further calculated as
\begin{subequations} \label{eq:rBP-5}
\begin{align}
    \eta_{i\to j} (\sigma_i) 
    &= \frac{1}{z_{i \to j}} \mathrm{e}^{\beta h_i \sigma_i - \frac{1}{2} \beta \lambda_i \sigma_i^2} \int \mathrm{d} \xi_{i \to j} ~ G(\xi_{i \to j}) \exp \left[ - \frac{\left( \xi_{i\to j} - M_{i \to j} \right)^2}{2 V_{i \to j}} \right] \\
    &= \frac{1}{z_{i \to j}} \mathrm{e}^{\beta h_i \sigma_i - \frac{1}{2} \beta \lambda_i \sigma_i^2} \int \mathrm{d} \xi_{i \to j} \exp \left[ - \frac{1}{2} \frac{1}{V_{i \to j}} \xi_{i \to j}^2 + \left( \frac{M_{i \to j}}{V_{i \to j}} + \sigma_i \right)  \xi_{i \to j} - \frac{M_{i \to j}^2}{2V_{i \to j}} \right]  \\
    &= \frac{1}{z_{i \to j}} \mathrm{e}^{\beta h_i \sigma_i - \frac{\beta \lambda}{2} \sigma_i^2} \exp \left( M_{i \to j} \sigma_i + \frac{1}{2} V_{i \to j} \sigma_i^2 \right) \\
    &= \frac{1}{z_{i \to j}} \exp \left[ - \frac{1}{2} \left( \beta \lambda_i - V_{i \to j} \right) \sigma_i^2 + \left( \beta h_i + M_{i \to j} \right) \sigma_i \right] . \label{eq:rBP-5d}
\end{align}
\end{subequations}
Equation~\eqref{eq:rBP-5d} implies that $\eta_{i\to j}(\sigma_i)$ follows a Gaussian distribution with the following mean and variance
\begin{align}
    & m_{i \to j} = \frac{\beta h_i + M_{i \to j}}{\beta \lambda_i - V_{i \to j}} , \\
    & v_{i \to j} = \frac{1}{\beta \lambda_i - V_{i \to j}} .
\end{align}

Now, we have obtained the so-called relaxed belief propagation (r-BP) equation:
\begin{equation} \label{eq:rBP-6}
    \left\{
        \begin{aligned}
            & M_{i \to j} = \beta \sum_{k\neq i, j} J_{ik} m_{k \to i} , \\
            & V_{i \to j} = \beta^2 \sum_{k\neq i, j} J^2_{ik} v_{k \to i} ,
        \end{aligned}
    \right. \qquad \qquad 
    \left\{
        \begin{aligned}
            & m_{i \to j} = \frac{\beta h_i + M_{i \to j}}{\beta \lambda_i - V_{i \to j}} , \\
            & v_{i \to j} = \frac{1}{\beta \lambda_i - V_{i \to j}} .
        \end{aligned}
    \right.
\end{equation}
Compared to the original belief propagation equation, the r-BP equations do not require a numerical integration, thereby improving the computational efficiency. Moreover, it provides a basis for deriving the AMP equation shown in the main text. After the relaxed belief propagation equation converges, we get further the marginal probability as
\begin{subequations} \label{eq:rBP-7}
\begin{align}
    \eta_{i} (\sigma_i) 
    &= \frac{1}{z_i} \mathrm{e}^{\beta h_i \sigma_i - \frac{1}{2} \beta \lambda_i \sigma_i^2} \prod_{j \neq i} \left\{ \int \mathrm{d}\sigma_j ~ \exp \left[ - \frac{1}{2} \left( \beta \lambda_j - V_{j\to i} \right) \sigma^2_j + \left( \beta h_j + \beta J_{ij} \sigma_i + M_{j \to i} \right) \sigma_j \right] \right\} \\
    &= \frac{1}{z_i} \mathrm{e}^{\beta  h_i \sigma_i - \frac{1}{2} \beta \lambda_i \sigma_i^2} \prod_{j \neq i} \exp \left[ \frac{\left( \beta h_j+ \beta J_{ij} \sigma_i + M_{j\to i} \right)^2}{2\left( \beta \lambda_j - V_{j \to i} \right)} \right] \\
    &= \frac{1}{z_i} \exp \Bigg[ - \frac{1}{2} \Big(\beta \lambda_i - \sum_{j \neq i} \beta^2 J_{ij}^2 v_{j \to i}\Big) \sigma_i^2 + \Big(\beta h_i + \sum_{j \neq i} \beta J_{ij} m_{j\to i}\Big) \sigma_i \Bigg]. \label{eq:rBP-7c}
\end{align}
\end{subequations}
As we expected, Eq.~\eqref{eq:rBP-7c} suggests that $\eta_{i}(\sigma_i)$ indeed follows a Gaussian distribution $\mathcal{N} (m_i, v_i)$, with the following mean and variance
\begin{equation}
    m_{i} = \frac{\beta h_i + \beta \sum_{j \neq i} J_{ij} m_{j \to i}}{\beta \lambda_i - \beta^2 \sum_{j \neq i} J^2_{ij} v_{j \to i}} , \qquad
    v_{i} = \frac{1}{\beta \lambda_i - \beta^2 \sum_{j \neq i} J^2_{ij} v_{j \to i}},
\end{equation}
where $m_{j \to i}$ and $v_{j \to i}$ are the fixed points of the r-BP iterative equations [Eq.~\eqref{eq:rBP-6}]. To conclude, the intuition of Gaussian approximation is theoretically justified.

\subsection{Approximate message passing equation}

If we can remove the directed arrows (e.g., $i\to j$) in the r-BP, the space complexity of the algorithm can be greatly reduced, reducing from $\mathcal{O}(D^4)$ to $\mathcal{O}(D^2)$. We next show how this can be possible. By using the following identities $M_i=M_{i\to j}+\beta J_{ij}m_{j\to i}$ and $V_i=V_{i\to j}+\beta^2J_{ij}^2v_{j\to i}$, we further find that 
\begin{equation}
\begin{aligned}
    m_{i} - m_{i \to j} = \frac{\beta h_i + M_{i}}{\beta \lambda - V_{i}} - \frac{\beta h_i + M_{i \to j}}{\beta \lambda - V_{i \to j}} = \frac{ \beta J_{ij} m_{j \to i} (\beta \lambda_i - V_i) + \beta^2 J^2_{ij} v_{j \to i} (\beta h_i + M_i) }{\beta^2 \lambda_i^2 - 2\beta \lambda_i V_i + V_{i}^2} \sim \mathcal{O} \left( \frac{1}{D^2} \right) ,
\end{aligned}
\end{equation}
and 
\begin{equation}
\begin{aligned}
    v_{i} - v_{i \to j} = \frac{1}{\beta \lambda - V_{i}} - \frac{1}{\beta \lambda - V_{i \to j}} = \frac{\beta^2 J^2_{ij} v_{j \to i}}{\beta^2 \lambda_i^2 - 2\beta \lambda_i V_i + V_{i}^2 } \sim \mathcal{O} \left( \frac{1}{D^4} \right).
\end{aligned}
\end{equation}
In the large $D$ limit (or even only $D$ is not too small), we can further simplify the relaxed belief propagation equation to the AMP equation
\begin{subequations}
\begin{align}
    m_{i} &= \frac{\beta h_i + \beta \sum_{j \neq i} J_{ij} m_{j}}{\beta \lambda_i - \beta^2 \sum_{j \neq i} J^2_{ij} v_{j}} , \\
    v_{i} &= \frac{1}{\beta \lambda_i - \beta^2 \sum_{j \neq i} J^2_{ij} v_{j}} .
    \end{align}
\end{subequations}
In this case, we only need to iterate the above two equations of $(m_i,v_i)$ to obtain the fixed points. This AMP equation saves $(D+1)^2$-fold space complexity for running the algorithm if we record $(D+1)^2$ as the number of spins in the system. In different forms, the AMP equation was first discovered in a statistical mechanics analysis of signal transmission problem~\cite{Kabashima-2003}, and is also rooted in the Thouless–Anderson–Palmer equation in glass physics~\cite{TAP-1977,Huang-2022,Mezard-1987}.

\section{Derivation of the unique optimal solution to the linear attention transformer}\label{secB}

In this section, we provide a detailed derivation of the unique optimal solution to the linear attention transformer (i.e., Eq. (10) and Eq. (11) in the main text).
 The input matrix $\mathbf{X}$ and the weight matrix $\mathbf{W}$ can be represented as $2\times2$ block matrices
\begin{equation}
    \mathbf{X} = 
    \left[\begin{array}{cc}
      \mathbf{X}_0 & \boldsymbol{\tilde{x}} \\
      \boldsymbol{y}_0^\top  & 0 
    \end{array}\right]
    , \quad
    \mathbf{W} = 
  \begin{bmatrix}
      \mathbf{W}_{11} & \mathbf{W}_{12} \\
      \mathbf{W}_{21} & \mathbf{W}_{22}
  \end{bmatrix} ,
\end{equation}
where $\mathbf{X}_0 = [\bm{x}_{1}, \bm{x}_{2}, \cdots, \bm{x}_{N} ] \in \mathbb{R}^{D\times N}$ denotes the data used for the prompts, $\bm{y}_0 = \mathbf{X}_0^\top \bm{w}/\sqrt{D}$ is the corresponding label vector, and $\boldsymbol{\tilde{x}}$ is the testing prompt whose label needs to be predicted by the transformer. The blocks of the weight matrix $\mathbf{W}$ have the same size as the corresponding blocks of $\mathbf{X}$, with $\mathbf{W}_{11} \in \mathbb{R}^{D\times D}$, $\mathbf{W}_{12} \in \mathbb{R}^{D\times 1}$, $\mathbf{W}_{21} \in \mathbb{R}^{1 \times D}$, and $\mathbf{W}_{22} \in \mathbb{R}$. Thus, we can calculate
\begin{equation}
    \mathbf{XX^\top} = 
    \left[\begin{array}{cc}
        \mathbf{X}_0 & \boldsymbol{\tilde{x}} \\
        \boldsymbol{y}_0^\top  & 0 
    \end{array}\right]
    \left[\begin{array}{cc}
        \mathbf{X}_0^\top & \boldsymbol{y}_0 \\
        \boldsymbol{\tilde{x}}^\top  & 0 
    \end{array}\right]
    = 
    \left[\begin{array}{cc}
        \mathbf{X}_0 \mathbf{X}_0^\top + \boldsymbol{\tilde{x}} \boldsymbol{\tilde{x}}^\top & \mathbf{X}_0 \boldsymbol{y}_0  \\
        \boldsymbol{y}_0^\top \mathbf{X}_0^\top  & \boldsymbol{y}_0^\top \boldsymbol{y}_0
    \end{array}\right] ,
\end{equation}
and
\begin{equation}
    \mathbf{WX} = 
    \left[\begin{array}{cc}
        \mathbf{W}_{11} & \mathbf{W}_{12} \\
        \mathbf{W}_{21} & \mathbf{W}_{22} 
    \end{array}\right] 
    \left[\begin{array}{cc}
        \mathbf{X}_0 & \boldsymbol{\tilde{x}} \\
        \boldsymbol{y}_0^\top  & 0
    \end{array}\right]
    = 
    \left[\begin{array}{cc}
        \mathbf{W}_{11} \mathbf{X}_0 + \mathbf{W}_{12}\boldsymbol{y}_0^\top & \mathbf{W}_{11} \boldsymbol{\tilde{x}} \\
        \mathbf{W}_{21} \mathbf{X}_0 + \mathbf{W}_{22} \boldsymbol{y}_0^\top & \mathbf{W}_{21} \boldsymbol{\tilde{x}}
    \end{array}\right] .
\end{equation}
Considering the prediction of the transformer $\hat{y} = \mathbf{Y}_{D+1,N+1} = \frac{1}{N}(\mathbf{XX^\top WX})_{D+1,N+1}$, using $2 \times 2$ block matrix multiplication, one obtains the following result:
\begin{subequations}
\begin{align}
    \hat{y} 
    & = \frac{1}{N} \Big[ \mathbf{XX^\top WX} \Big]_{D+1, N+1}  \label{eq:pred-1} \\
    & = \frac{1}{N} \Big[ 
        \boldsymbol{y}_0^\top \mathbf{X}_0^\top \mathbf{W}_{11} \tilde{\boldsymbol{x}} + \boldsymbol{y}_0^\top \boldsymbol{y}_0 \mathbf{W}_{21} \tilde{\boldsymbol{x}}
    \Big]  \label{eq:pred-2} \\
    & = \frac{1}{N} \bigg[ 
        \left( \frac{1}{\sqrt{D}} \mathbf{X}_0^\top \boldsymbol{w}_\star \right)^\top \mathbf{X}_0^\top \mathbf{W}_{11} \tilde{\boldsymbol{x}} + \left( \frac{1}{\sqrt{D}} \mathbf{X}_0^\top \boldsymbol{w}_\star \right)^\top \left( \frac{1}{\sqrt{D}} \mathbf{X}_0^\top \boldsymbol{w}_\star \right) \mathbf{W}_{21} \tilde{\boldsymbol{x}}
    \bigg]  \label{eq:pred-3} \\
    & = \frac{1}{N} \bigg[ 
        \frac{1}{\sqrt{D}} \boldsymbol{w}_\star^\top \mathbf{X}_0 \mathbf{X}_0^\top \mathbf{W}_{11} \tilde{\boldsymbol{x}} + \frac{1}{D} \boldsymbol{w}_\star^\top \mathbf{X}_0 \mathbf{X}_0^\top \boldsymbol{w}_\star \mathbf{W}_{21} \tilde{\boldsymbol{x}}
    \bigg]  \label{eq:pred-4} \\
    & = \frac{1}{\sqrt{D}} \boldsymbol{w}_\star^\top \mathbf{C}_{0} \bigg( \mathbf{W}_{11} + \frac{1}{\sqrt{D}} \boldsymbol{w}_\star \mathbf{W}_{21} \bigg) \tilde{\boldsymbol{x}},
    \label{eq:pred-5}
\end{align}
\end{subequations}
where $\boldsymbol{w}_\star$ is the task vector, the definition $\boldsymbol{y}_0 = \mathbf{X}_0^\top \boldsymbol{w} _\star/ \sqrt{D}$ is used to obtain Eq.~\eqref{eq:pred-3}, and the new definition $\mathbf{C}_{0} \equiv \frac{1}{N} \mathbf{X}_0 \mathbf{X}_0^\top$ is used in Eq.~\eqref{eq:pred-5}. Comparing the actual prediction with the true label $\tilde{\bm{y}} = \bm{w}_\star^\top \bm{\tilde{x}} / \sqrt{D}$ corresponding to $\bm{\tilde{x}}$, one finds that the optimal solution for the weight matrix must satisfy the following condition:
\begin{equation} \label{eq:optimal}
  \mathbf{C}_{0} \bigg( \mathbf{W}_{11} + \frac{1}{\sqrt{D}} \boldsymbol{w}_\star \mathbf{W}_{21} \bigg) = \id_D .
\end{equation}

In the asymptotic limit, i.e., $N\propto D$, $C_0$ is a Wishart random matrix. However, in the non-asymptotic case we consider, $D$ is kept finite while $N$ can be independently large, we get an identity matrix of $\mathbf{C}_0$, which can be proved by using the central limit theorem. Therefore, Eq.~\eqref{eq:optimal} can be simplified as
\begin{equation} \label{eq:optimal-2}
    \mathbf{W}_{11} + \frac{1}{\sqrt{D}} \boldsymbol{w}_\star \mathbf{W}_{21} = \id_D .
\end{equation} 
 
Equation~\eqref{eq:optimal-2} is actually a set of linear equations, which can be written in the form of $\mathbf{A} \bm{x} = \bm{b}$ where $\{x_i\}$ indicates the weight components to be determined.  The number of variables in the equations is $D^2 + D$. Given an input matrix $\mathbf{X}$ generated by a specific $\bm{w}$, we can write down $D^2$ equations to solve for the optimal weight. Therefore, the mathematics of linear equations tells us that when the number of input matrices $P=1$, the weight matrix $\mathbf{W}$ has infinitely many solutions. However, as long as $P > 1$, the set of equations has a unique solution, which can be readily deduced as $\mathbf{W}_{11}^\star = \id_D$ and $\mathbf{W}_{21}^\star = \bm{0}$. In practice, the transformer or our spin glass model needs a larger value of $P$ to identify this optimal matrix.

\section{Convexity and Hessian matrix of the linear attention model} 

In this section, we discuss the convexity of our model used in the main text by analyzing the spectral density of the Hessian matrix.

\begin{figure}[h]
  \centering
  \includegraphics{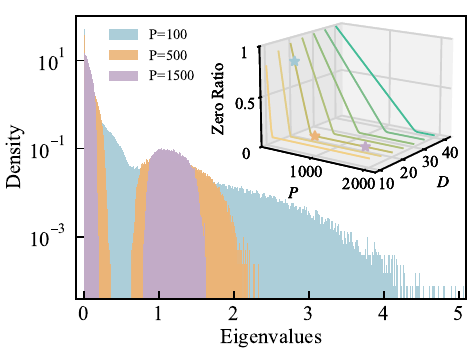}
  \caption{
    The spectral density of the Hessian matrix $\mathbf{H}$ for $D=20$, $N=100$, and different values of $P$. The results are obtained from the statistics of $5\,000$ matrices. 
    The inset in the upper right corner shows the proportion of zero modes for different values of $P$ and $D$. Stars in the inset correspond to the cases plotted in
   in the main plot with the corresponding color.
  } 
  \label{fig-S5}
\end{figure}

 After the spin mapping $\hat{y}^\mu = \sum_{i} \sigma_i s_i^\mu$, the loss function is written as $\mathcal{L} = (2P)^{-1} \sum_{\mu=1}^{P} (y^\mu - \hat{y}^\mu)^2$, where $s_i^\mu$ is a vectorization of the matrix $\mathbf{S}^\mu$. The Hessian matrix can be calculated as~\cite{PAI-2025}
\begin{equation}
  \mathbf{H}_{ij} = \frac{\partial^2 \mathcal{L}}{\partial \sigma_i \partial \sigma_j} = \frac{1}{P} \sum_{\mu=1}^{P} s_i^\mu s_j^\mu=\frac{1}{P}[\mathbb{S}^\top\mathbb{S}]_{ij},
\end{equation}
where $\mathbb{S}\in\mathbb{R}^{P\times (D+1)^2}$. It can be thus proved that $\mathbf{H}$ is a positive semidefinite matrix.
The spectral density of the Hessian matrix is shown in Fig.~\ref{fig-S5}. The eigenvalues are all non-negative, which indicates that the Hessian matrix is positive semidefinite. This means that the loss function is (but not strictly) convex, and we find that the task diversity affects the number of zero modes in the energy landscape (Fig.~\ref{fig-S5}). The large number of zero eigenvalues implies that the energy landscape of the model is marginally stable in many local regions, resulting in a high-dimensional landscape of many minima [see Fig.~3\,(a) for the case of $P=10$], which reflects the algorithmic hardness when $P$ is small. Interestingly, this feature is also observed in Boltzmann machine learning~\cite{Remi-2012}.

\section{Generalization to more complex attention structures}

In this section, we verify whether our spin glass model is robust to more complex attention structures in transformers, by carrying out extensive experimental simulations.
  The linear attention setting used in the main text is called simplified linear attention (SLA) in this section. The explicit formulation is given below.
\begin{equation} \label{eq:SLA}
    \mathbf{Y}_{\rm SLA} = \frac{1}{N} \mathbf{XX^\top WX},
\end{equation}
where $\mathbf{W} \equiv \mathbf{W}_{\rm Q}^\top \mathbf{W}_{\rm K}$.
This SLA will be compared with the case of removing the softmax function yet keeping the value matrix trainable, which we call the full linear attention (FLA) as follows. 
\begin{equation}\label{eq:FLA}
    \mathbf{Y}_{\rm FLA} = \frac{1}{N} \mathbf{W}_{\rm V} \mathbf{XX^\top} \mathbf{W} \mathbf{X}.
\end{equation}
The single-head full softmax attention (SA) defined below is also compared.
\begin{equation} \label{eq:SA}
    \mathbf{Y}_{\rm SA} = \mathbf{W}_{\rm V} \mathbf{X} \cdot \mathrm{Softmax} \left( \frac{(\mathbf{W}_{\rm Q} \mathbf{X})^\top \mathbf{W}_{\rm K} \mathbf{X}}{\sqrt{H}} \right),
\end{equation}
where $\mathbf{W}_Q, \mathbf{W}_K \in \mathbb{R}^{H \times (D+1)}$ and $\mathbf{W}_V \in \mathbb{R}^{(D+1) \times (D+1)}$, and $H$ is the internal size of the attention operation and can be equal to the embedding dimension.
We also consider a small full transformer encoder, with a feed-forward layer, residual connection, and linear readout head, defined as
\begin{equation} \label{eq:TF}
  \mathbf{Y}_{\rm TF} = \mathbf{z}_{1} + \mathbf{z}_{3},
\end{equation}
with
\begin{subequations}
\begin{align}
  & \mathbf{z}_{1} = \mathrm{SA}(\mathbf{X}) + \mathbf{X}, \\
  & \mathbf{z}_{2} = \mathrm{ReLU} \left( \mathbf{W}_1 \cdot \mathbf{z}_{1} + b_1 \right), \\
  & \mathbf{z}_{3} = \mathbf{W}_2 \cdot \mathbf{z}_{2} + b_2 ,
\end{align}
\end{subequations}
where $\mathrm{SA}(\mathbf{X})$ is the standard self-attention layer as Eq.~\eqref{eq:SA}, and $\mathbf{W}_1, \mathbf{W}_2$ and $b_1, b_2$ are weights and biases of the two feed-forward layers, respectively. The weight matrices in $\mathrm{SA}(\mathbf{X})$ are $\mathbf{W}_Q, \mathbf{W}_K \in \mathbb{R}^{H \times (D+1)}$ and $\mathbf{W}_V \in \mathbb{R}^{(D+1) \times (D+1)}$, respectively, and $H$ is the internal size of the attention operation. 
We call this complex layered transformer structure TF.

In addition, the transformer usually employs the multi-head attention structure. Therefore, for comparison, we further consider a multi-head linear attention model (MHLA), where the multiple outputs of Eq.~\eqref{eq:FLA} (one output corresponds to one head) are concatenated and then linearly read through a learnable readout matrix $\mathbf{W}_O$. The formulation is given by $\mathbf{Y}_{\rm MHLA}=\sum_{h=1}^{M}\mathbf{W}_O^{h}\mathbf{Y}_{{\rm FLA}}^h$. We denote the number of attention heads, i.e., the dimension of the linear readout, as $M$.

\begin{figure}[h]
  \centering
  \includegraphics{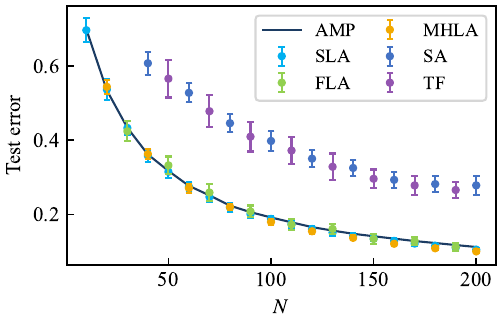}
  \caption{Comparison of generalization errors for different transformer settings. $\lambda=0.01$, $P=10\,000$, $D=20$, and $\beta=100$ for AMP.  For SA and TF, $H=32$. For MHLA, $M=10$. Results are averaged over $100$ independent trials. Symbols indicate SGD results.
  } 
  \label{fig-S6}
\end{figure}

 The experimental results of the above different models are shown in Fig.~\ref{fig-S6}. Figure~\ref{fig-S6} shows the robustness of our theory against different transformer settings, although SA and TF display a higher test error. This can be understood as the linear attention structure is more suitable for the linear regression task compared to the non-linear softmax attention. For the softmax attention and more advanced structures, it is hard to derive such concise results as in Sec~\ref{secB}, and the spherical spin glass model specified in the main text. However, the ground state interpretation of the learning is not a specific picture, and the underlying mechanism for the ICL in the simple but non-trivial setting we consider can be analytically clarified, thereby offering a promising avenue for thinking about how to model many intriguing but puzzling properties of large language models. 
 
 We leave the exploration of more complex tasks beyond linear regression and more 
complex transformers such repeated transformer blocks together with multi-headed structures to future works. The current experiments are sufficient to justify our theory of spin-glass mapping.


\end{document}